\documentclass[sn-mathphys,Numbered]{sn-jnl}

\usepackage{graphicx}%
\usepackage{multirow}%
\usepackage{amsmath,amssymb,amsfonts}%
\usepackage{amsthm}%
\usepackage{mathrsfs}%
\usepackage[title]{appendix}%
\usepackage{xcolor}%
\usepackage{textcomp}%
\usepackage{manyfoot}%
\usepackage{booktabs}%
\usepackage{algorithm}%
\usepackage{algorithmicx}%
\usepackage{algpseudocode}%
\usepackage{listings}%
\usepackage{lineno}%

\begin{document}

\title{High-energy tunable ultraviolet pulses generated by optical leaky wave in filamentation}


\author[ ]{\fnm{Litong} \sur{Xu}}

\author*[ ]{\fnm{Tingting} \sur{Xi}}\email{ttxi@ucas.ac.cn}

\affil[ ]{\orgdiv{School of Physical Sciences}, \orgname{University of Chinese Academy of Sciences}, \city{Beijing}, \postcode{100049}, \country{China}}


\abstract{Ultraviolet pulses could open up new opportunities for the study of strong-field physics and ultrafast science. However, the existing methods for generating ultraviolet pulses face difficulties in fulfilling the twofold requirements of high energy and wavelength tunability simultaneously.  Here, we theoretically demonstrate the generation of high-energy and wavelength tunable ultraviolet pulses in preformed air-plasma channels via the leaky wave emission. The output ultraviolet pulse has a tunable wavelength ranging from 250 nm to 430 nm and an energy level up to sub-mJ. An octave-spanning ultraviolet supercontinuum with a flatness better than 3 dB can also be obtained via longitudinally modulated dispersion. Such a high-energy tunable ultraviolet light source may provide promising opportunities for characterization of ultrafast phenomena such as molecular breakup, and also an important driving source for the generation of high-energy attosecond pulses.
}




\maketitle

 \section{Introduction}
 
 Ultraviolet (UV) pulses have gained increasing popularity in the last few decades due to their versatile applications in biology \cite{matsumoto2019}, material fabrication \cite{lombardo2019} and spectroscopy \cite{mao2021}. With the rapid development of laser technology, UV pulses in femtosecond or even attosecond time scales have been generated and widely applied \cite{xie2024,geneaux2016}. Within the entire UV range, the UV pulses with wavelengths from 200 nm to 400 nm remain highly required because they can be resonantly absorbed by most substances and can serve as unique light sources to study the ultrafast dynamics of atoms and molecules, for instance, measuring absolute transition frequency of atoms \cite{ozawa2013}, revealing structure and dynamics of biomolecules \cite{borrego-varillas2019}, and characterizing molecular breakup \cite{yue2015}. Moreover, when the UV pulse with high energy and tunable wavelength is used as a single driving source or one of the two-color fields, it enables the generation of high-order harmonics with high conversion efficiency \cite{marceau2017,midorikawa2022}, which will be beneficial for the applications that require high photon flux, such as attosecond spectroscopy \cite{krausz2009} and extreme-UV lithography \cite{tseng2023}.

 At present, UV pulses in the spectral region of 200 nm to 400 nm are mainly generated from low-order harmonics \cite{zhou2014} and four-wave mixing \cite{lekosiotis2020}. Nevertheless, the wavelength of the generated UV pulse is often limited by the available wavelength of the pump pulse. To overcome this problem, UV pulses with continuously tunable wavelength can be generated from resonant dispersive wave in gas-filled hollow-core fibers by varying the gas pressure \cite{saleh2011,mak2013}. In this case, the energy of UV pulses has reached a few tens of $\mathrm{\mu}$J, but it is difficult to be further enhanced due to the limited length and diameter of the fiber and the available energy of the few-cycle pump pulses \cite{travers2019}. Although the removal of the fiber confinement could break the restriction on energy enhancement, it also leads to the lack of negative dispersion provided by fiber waveguides. Correspondingly, UV pulses cannot be obtained because the propagation of the near-infrared femtosecond laser pulse in uniform gas results in supercontinuum emission with cut-off wavelength of $\sim 400$ nm. 
 
 In this article, we propose to generate wavelength-tunable UV pulses from the supercontinuum emission of the near-infrared femtosecond pulse by introducing a preformed air-plasma channel with the adjusted negative dispersion into air. The output UV pulse has a tunable wavelength ranging from 250 nm to 430 nm and an energy level up to sub-mJ, suggesting that this approach holds great potential in energy scaling of the tunable UV pulses. Besides, an octave-spanning UV supercontinuum with a flatness better than 3 dB can be obtained via the longitudinally modulated dispersion, which should be of particular interest to ultrafast spectroscopy \cite{hong2023} and attosecond physics \cite{krausz2009}.

  \begin{figure}[htbp]\centering
 \includegraphics[width=0.8\textwidth]{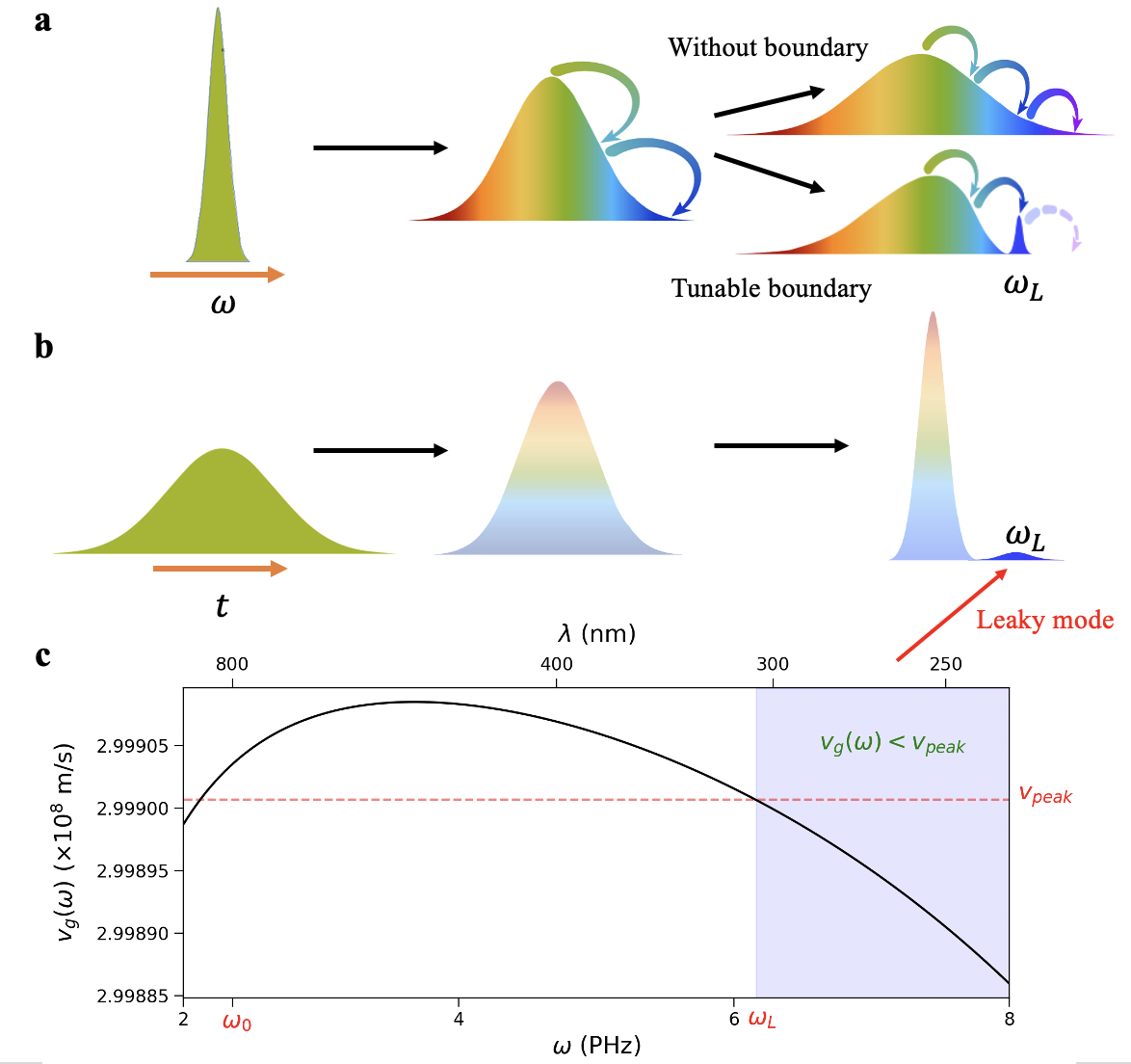}%
 \caption{\label{fig1}\textbf{Schematic illustration of leaky wave generation.} (a) The spectral broadening process of filamentation when there is no spectral boundary (top) and a spectral boundary (bottom). (b) The pulse self-compression and leaky wave generation in the tail of the pulse. (c) The group velocity $v_g$ as a function of frequency $\omega$ when the plasma density is $\rho_0=1.5\times 10^{17}$ cm$^{-3}$ . The red dashed line denotes the velocity of pulse peak $v_{peak}$. The colored area represents the leaky mode, with $v_g(\omega)<v_{peak}$.}
 \end{figure}

\section{Conception}
Our basic idea is illustrated in Fig. \ref{fig1}. During filamentation in air, spectral broadening induced by self-phase modulation can be viewed as a cascaded process.
The spectral energy is transferred from the central frequency to high frequency components, and then higher. This process could lead to supercontinuum generation with blue-side cut-off wavelength $\sim 400$ nm, which is the common situation for air filamentation. The spectrum in UV region cannot be obtained because air is normally dispersive, and the high-frequency components move far behind the pulse peak gradually, with their intensity too low to participate in further frequency shifting. To overcome this limitation, we introduce a preformed uniform plasma channel, which can introduce negative dispersion into air. Based on this negative dispersion, the high-frequency components can be constrained near the peak of the pulse. They can undergo further frequency shift to the UV region.
As shown in Fig. \ref{fig1}a, the cascaded frequency shift combined with the negative dispersion could result in the super-broad spectrum, which covers the UV region. To obtain an isolated UV spectral hump, we need a frequency boundary $\omega_L$ to terminate the cascaded process. For the components with frequency higher than $\omega_L$, they fall behind the primary pulse and will not undergo further frequency shift due to the relatively low intensity. Therefore, the spectral energy is deposited in this region, as shown in Fig. \ref{fig1}. We term this process temporal leaky wave emission, because it is similar to the spatial leaky wave in the waveguide \cite{monticone2015}.

We will show the frequency boundary $\omega_L$ of the cascaded broadening, which corresponds to the peak frequency of the leaky UV spectrum, could be tuned by adjusting the density of the preformed plasma channel. To calculate $\omega_L$, we need to find the spectral components which fall behind the primary pulse. Here we use a preformed air-plasma channel with plasma density $\rho_0$ to provide tunable dispersion condition, and the dispersion relation is given by

\begin{equation}
k(\omega)=\frac{n_a(\omega)\omega}{c}-\frac{\rho_0\omega}{2c\rho_c},
\end{equation}
where $n_a(\omega)$ is the refractive index of air,  and $\rho_{c}=\epsilon_0m_e\omega^2/e^2$ is the frequency dependent critical plasma density that accounts for plasma dispersion.
For the velocity of the primary pulse, we use the group velocity of the pulse peak with central frequency $\omega_0$ and consider the self-steepening effect. It is written as
\begin{equation}
v_{peak}=\frac{1}{k'(\omega_0)+n_2I_p/c},
\end{equation}
where $n_2$ is the nonlinear refractive index coefficient and $I_p$ is the peak intensity of the pulse. The group velocity of spectral components falling behind the primary pulse is given by $v_g(\omega)<v_{peak}$, and the critical frequency $\omega_L$ satisfies 

\begin{equation}\label{eq-pre}
v_g(\omega_L)=v_{peak}.
\end{equation}
From the above analysis, we can see that the peak frequency of the UV spectrum $\omega_L$ could be tuned by adjusting the plasma density $\rho_0$.

To verify that the dispersion induced by the preformed low-density plasma channel can indeed support the generation of UV pulses via temporal leaky-wave emission, we calculate the group velocity of different spectral components for the plasma density $\rho_0=1.5\times 10^{17}$ cm$^{-3}$, $n_2=0.96\times 10^{-19}\ \mathrm{cm^2/W}$ and $I_p=100$ TW/cm$^2$, as shown in Fig. \ref{fig1}c. For the blueshift components with frequency smaller than  $\omega_L$, their group velocities are larger than the pulse peak. Although they are generated at the pulse trailing edge by self-phase modulation, they can be confined within the primary pulse.
For spectral components with frequency larger than $\omega_L$, their group velocities are lower than the pulse peak. Therefore, this part of the spectrum will lag behind the primary pulse. This evolution of group velocity as a function of the frequency supports our scheme of the temporal leaky-wave emission. For the plasma density $\rho_0=1.5\times 10^{17}$ cm$^{-3}$, the corresponding wavelength of the leaky-wave emission is 320 nm, located in the UV region.

To confirm our model, we numerically solve the UPPE (see Methods) to study the propagation of femtosecond laser pulses in air-plasma channels. The input 800 nm pulse has a duration of 30 fs, a beam waist of 2 mm and a peak power of $P_{\mathrm{in}}=5P_{\mathrm{cr}}$. It is focused by a lens with $f=1$ m.

\section{Discussion of results}
\subsection{Tunable UV pulse}

\begin{figure}[htbp]\centering
 \includegraphics[width=\textwidth]{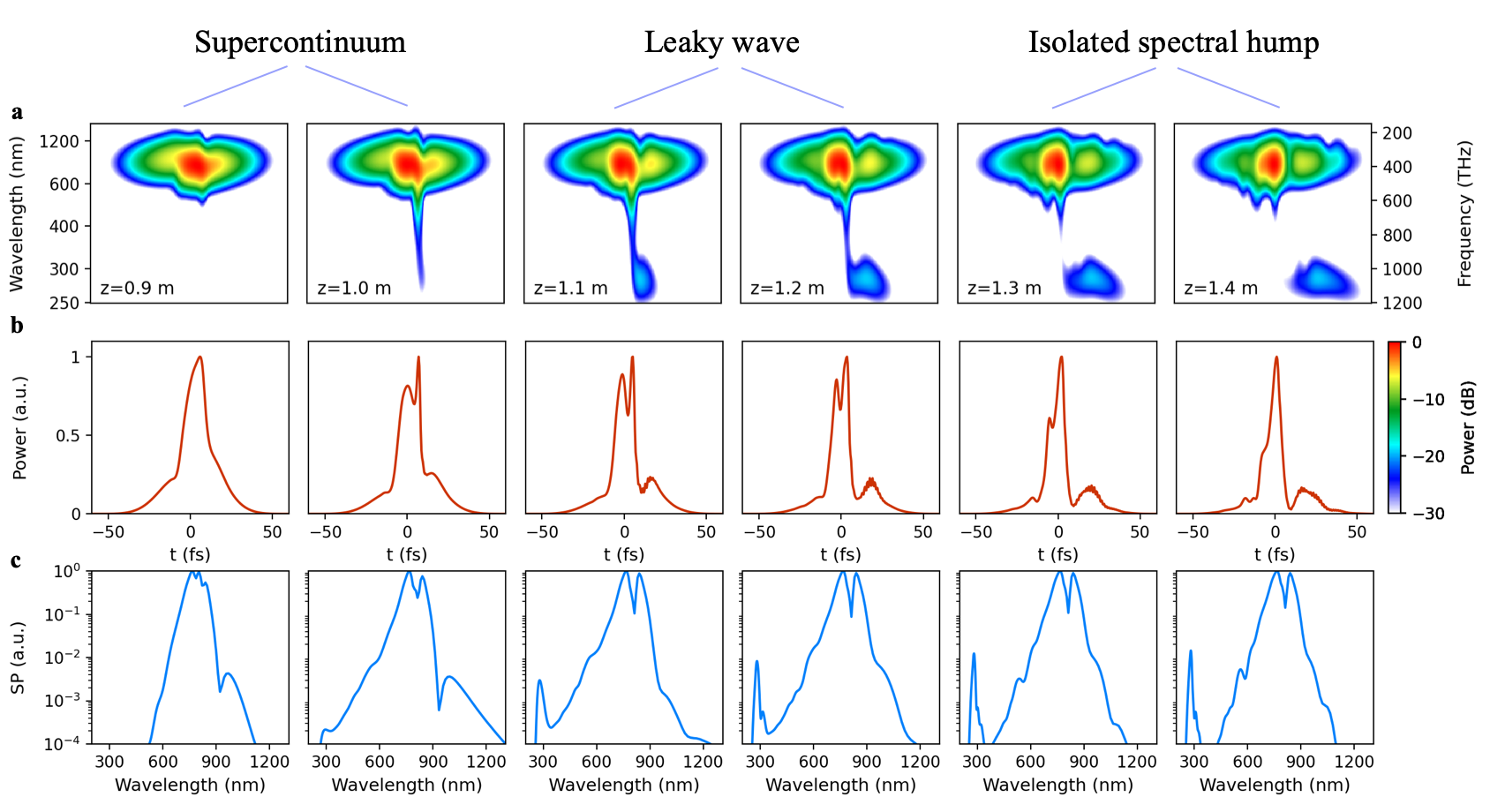}%
 \caption{\label{fig2}\textbf{Schematic of leaky wave generation in a preformed air-plasma channel}. (a) Spectrogram, (b) pulse profile and (c) spectral power (SP) distribution at different positions. The density of the air-plasma channel is  $\rho_0=1.5\times 10^{17}$ cm$^{-3}$.}
 \end{figure}

Firstly, we investigate the pulse dynamics when the density of preformed air-plasma channel $\rho_0=1.5\times 10^{17}$ cm$^{-3}$. Filamentation can be divided into three stages. At z=1 m, we can see supercontinuum generation (Fig. \ref{fig2}a) as well as pulse compression (Fig. \ref{fig2}b). Pulse-splitting also occurs at this stage (Fig. \ref{fig2}b), and it can be seen that the supercontinuum is mainly contributed by the rear sub-pulse. Since the supercontinuum has reached the critical wavelength of leaky wave, at $z=1.1$ m there arises a high-frequency lobe in the spectrogram (see Methods), which corresponds to the newly generated sub-pulse in Fig. \ref{fig2}b and the spectral hump at 283 nm in Fig. \ref{fig2}c. Then the leaky wave is continuously generated and lagged behind, which makes the time span and the spectral intensity of leaky wave increase. As the spectral components between $\omega_0$ and $\omega_L$ are consumed , we can see the formation of isolated spectral hump. The simulation results confirm our scheme, except for a slight difference in the peak wavelength of the UV hump. The reason for this difference will be discussed below.

  \begin{figure}[htbp]\centering
 \includegraphics[width=0.6\textwidth]{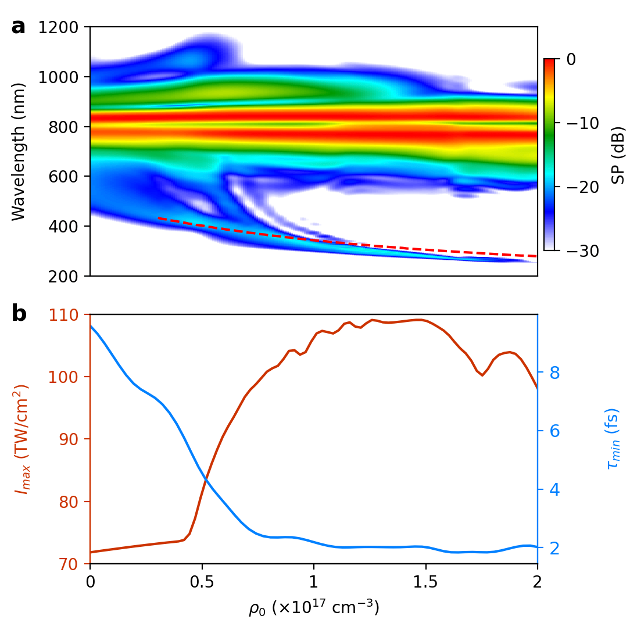}
 \caption{\label{fig3}\textbf{Frequency tunning via changing density of air-plasma channels.} (a) Output spectral power (SP) distribution  as $\rho_0$ changes from 0 to $2\times10^{17}$ cm$^{-3}$. The red dashed line denotes the central wavelength of leaky wave predicted by Eq. (3). (b) Maximum intensity (red)  and minimum on-axis pulse duration (blue) during propagation for different values of $\rho_0$.}
 \end{figure}

For different preformed plasma densities $\rho_0$, the output spectra are shown in Fig. \ref{fig3}a. When $\rho_0<0.3\times 10^{17}$ cm$^{-3}$, the spectrum is broadened without isolated UV spectral hump. As the density of the pre-plasma increases, the stronger negative dispersion leads to pulse self-compression, resulting in a higher peak intensity and a shorter pulse duration, as shown in the Fig. \ref{fig3}b.
This gives rise to a much steeper intensity slope $\partial_t I$, thus a much broader spectrum, which covers the critical wavelength of leaky wave $\lambda_L=2\pi c/\omega_L$  and an isolated UV spectral hump is formed. The central wavelength of the UV spectral hump decreases as $\rho_0$ increases, resulting in a wavelength tuning range between 250 nm and 430 nm. The simulation results accord well with the theoretically predicted value of $\lambda_L$ from Eq. (3), denoted by the red dashed line in Fig. \ref{fig3}(a). The deviation between our model and simulation mainly comes from the post-frequency-shift process of leaky wave: the spectral component with wavelength of $\lambda_L$   generated near the pulse peak undergoes further frequency shift as it moves toward the pulse tail. The lower bound of wavelength tuning in our scheme is about 250 nm, which is limited by two factors. For one thing, the conversion efficiency decreases with the output wavelength. For another, the clamping intensity limits the spectral width of the supercontinuum, while to generate leaky wave, the supercontinuum has to cover the critical wavelength. The above restriction may be broken by using a pump beam with shorter central wavelength.

\subsection{Energy scaling}
  \begin{figure}[htbp]\centering
 \includegraphics[width=\textwidth]{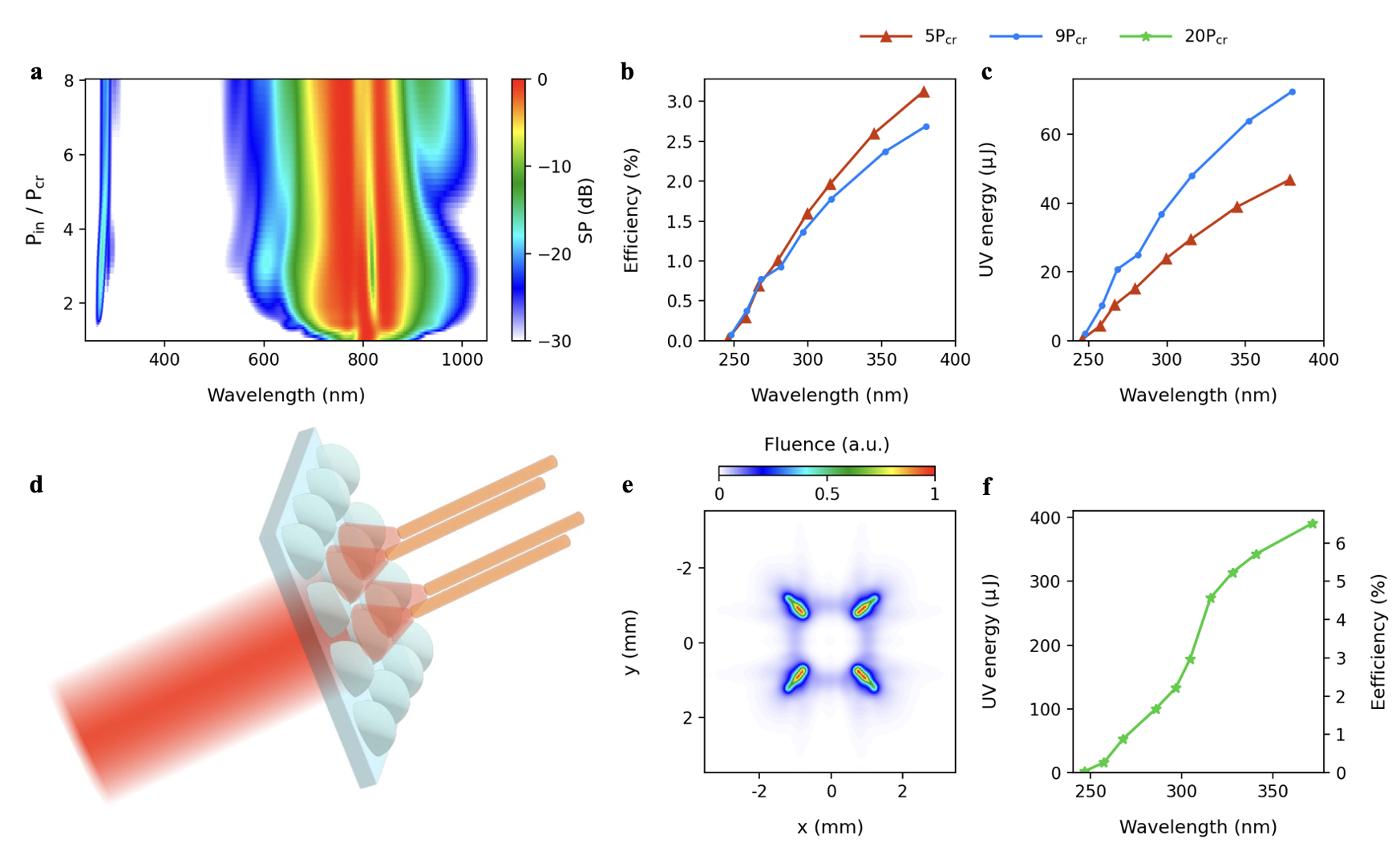}%
 \caption{\label{fig4}\textbf{Energy scaling proposals of the UV radiation.} (a) Output spectral power (SP) distribution as varying the input power. (b) The conversion efficiency and (c) the energy of the output UV pulses for different input power and output wavelength. (d) Schematic illustration of regularization of multiple filaments with a microlens array. (e) The fluence distribution and (f) the conversion efficiency and output UV pulse energy when using a microlens array.}
 \end{figure} 
 
 As an important criteria of UV light source, the energy of UV pulse is typically a few $\mathrm{\mu J}$ using photonic crystal fibers and hollow-core capillary fibers. The fiber system, though providing a simple and flexible platform to generate tunable UV pulse source, also puts an intrinsic limitation on the available energy. Our air-plasma channel scheme could be a promising candidate for obtaining unprecedented high energy UV pulse, without the concern of damage to the media. To explore the energy scaling rules of our scheme, we first investigate the influence of input pulse peak power on the output spectra. We fix $\rho_0$ at $1.5\times 10^{17}$ cm$^{-3}$ and gradually increase the input power from $P_{\mathrm{cr}}$ to $9P_{\mathrm{cr}}$. The output spectra evolution is shown in Fig \ref{fig4}a. Leaky wave emission with a wavelength of  280 nm is generated when the input peak power $P_{\mathrm{in}}\approx 1.5P_{\mathrm{cr}}$. When increasing the input power, the leaky wavelength first becomes longer (about 10 nm), because the minimum pulse width is slightly shorter, and the post-frequency-shift effect mentioned before is weaker. 
 For $P_{\mathrm{in}}> 5P_{\mathrm{cr}}$, the minimum pulse width as well as the leaky wavelength is almost constant, which ensures the stability of our scheme. As the incident power increases, the  UV spectral hump becomes more intense with an increased energy and a slightly decreased conversion efficiency. For $P_{\mathrm{in}}= 5P_{\mathrm{cr}}$  and $9P_{\mathrm{cr}}$, the UV spectrum has a central wavelength of 283 nm and 285 nm, an energy of 13.3 $\mathrm{\mu J}$ and 22.4 $\mathrm{\mu J}$,  and a conversion efficiency of 0.88\% and 0.83\%, respectively.
 We also calculate the energy and conversion efficiency of the UV spectrum with different central wavelength by adjusting pre-plasma density. As expected, the conversion efficiency is lower for shorter wavelength due to the cascaded feature of spectral broadening, as shown in Fig. \ref{fig4}b. For wavelength shorter than 280 nm, the conversion efficiency is lower than 1\%. However, since the conversion efficiency only decreases slightly when the input peak power is increased from $5P_{\mathrm{cr}}$ to $9P_{\mathrm{cr}}$ (less than 0.5\%), energy scaling of UV pulses could be directly achieved by increasing the input power, as indicated by Fig. \ref{fig4}c. The output energy of UV pulses could be  more than 40 $\mathrm{\mu J}$ for wavelength longer than 300 nm, which is several times the value in literature \cite{travers2019}.

If further increasing the input power, the influence of multiple filaments should be considered. Here we propose that a microlens array (MLA) may be used to obtain regularly distributed filaments, as demonstrated in Fig. \ref{fig4}d. Since the supercontinuum emitted from each single filament can be  coherently combined \cite{qian2021b}, using a MLA to organize the filaments generated by high-energy pulse could be a promising solution to mJ level UV pulse. As a proof-of-principle simulation, we use an incident pulse with peak power $P_{\mathrm{in}}= 20P_{\mathrm{cr}}$, focused by a MLA (the focal length and size of each square lenslet are 1 m and 2 mm, respectively). Four filaments are formed (Fig. \ref{fig4}e), and the output UV pulse energy reaches sub-mJ level, as shown in Fig. \ref{fig4}f. Note that when using MLA, the conversion efficiency is nearly doubled, about 6\% for a 350 nm pulse. This is related to the interaction of the energy background of filaments: the energy that is defocused by one filament can be reused by other filaments. It can also be seen that it is necessary to provide a uniform plasma channel with a centimeter-diameter for the generation of high-energy UV pulses. Such a plasma channel has been reported to be generated by filamentation of high-energy picosecond laser pulses \cite{tochitsky2019}.

\begin{figure}[htbp]\centering
 \includegraphics[width=\textwidth]{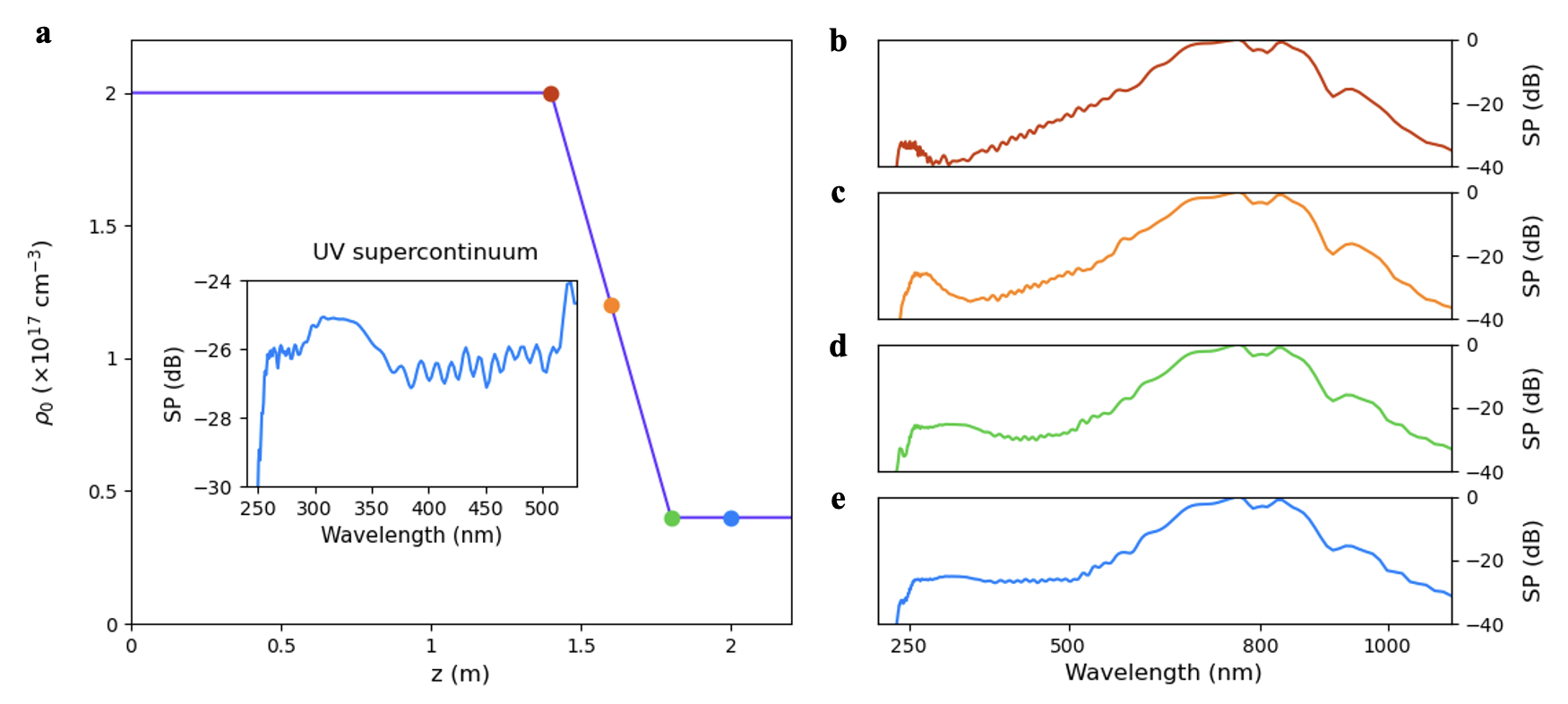}%
 \caption{\label{fig5}\textbf{Generation of superflat UV continuum in an air-plasma channel with density gradient.} (a) Density distribution of the preformed air-plasma channel. (b-e) Spectral power (SP) distribution at the four positions  marked in (a). Inset in (a) gives a detailed view of the output UV supercontinuum.}
 \end{figure}

\subsection{Superflat UV supercontinuum}

From Fig. \ref{fig3}a we know wavelength tuning of UV pulse can be achieved by changing the plasma density. To generate a flat UV spectrum, we could use a pre-plasma with a density gradient along the propagation direction. We design a plasma gradient shown in Fig. \ref{fig5}a, where the density of pre-plasma remains constant at $2\times 10^{17}$ cm$^{-3}$ initially, then decreases linearly from $z=1.4$ m to $z=1.8$ m, and finally remains the density of $0.4\times 10^{17}$ cm$^{-3}$. For an input laser power $P_{\mathrm{in}}= 9P_{\mathrm{cr}}$, the spectra at different positions are given in Fig. \ref{fig5}b-e. At $z=1.4$ m, a UV spectral hump is formed near 250 nm, which is consistent with the leaky wavelength for  $\rho_0=2\times 10^{17}$ cm$^{-3}$  shown in Fig. \ref{fig3}a. As the density of the pre-plasma decreases, the corresponding leaky wavelength decreases and the spectral gap between 250 nm and 500 nm is gradually filled. As a result, at $z=2$ m we obtain a supercontinuum that spans from 250 nm to 1100 nm, more than two octaves. Moreover, we find that the supercontinuum in the UV region is superflat, with very close spectral power. The UV supercontinuum is detailed in the inset of Fig. \ref{fig5}a, where the spectral power fluctuates less than 3 dB between 260 nm and 520 nm. The flatness of the UV supercontinuum  ensures consistent measurement quality throughout the spectrum of interest, which can be particularly useful to increase the signal-to-noise ratio in ultrafast spectroscopy \cite{hong2023}. The ultrabroad and superflat UV spectrum is also highly beneficial for generating UV few-cycle pulses, serving as important driving sources for high-order harmonic generation \cite{krausz2009}.

\section{Conclusion}
In summary, we have proposed the leaky wave emission model that enables the generation of wavelength-tunable ultraviolet pulses. The model was verified by using a pre-formed plasma channel to modulate the dispersion condition of air in simulation. Compared to the resonant dispersive wave scheme in hollow-core fibers, our approach removes the limitation of fiber system, thus showing great potential in energy scaling. The output UV pulse has a tunable wavelength range of 250 nm to 430 nm and an energy level up to sub-mJ. When applying a longitudinally modulated plasma density, we could obtain an octave-spanning ultraviolet supercontinuum with a flatness better than 3 dB, which could greatly facilitate the applications in ultrafast spectroscopy. This study could provide an essential light source for ultrafast science and important driving source for attosecond physics.

\section{Methods}
\textbf{UPPE model}
The unidirectional pulse propagation equation (UPPE) that governs the forward-propagating electric field envelope $\hat{E}(k_x,k_y,\omega,z)=\mathcal{F}[E(x,y,t,z)]$ in the pulse local frame ($t\rightarrow t-z/v_g$) is written as \cite{couairon2011}:
\begin{equation}
\partial_z\hat{E}=i\left(K_z-k_0-\frac{\omega-\omega_0}{v_g}\right)\hat{E}+\frac{i\omega^2}{2\epsilon_0c^2K_z}\mathcal{F}[P_{NL}+\frac{i}{\omega}J].
\end{equation}
Here $K_z=\sqrt{k(\omega)^2-k_x^2-k_y^2}$, $k(\omega)=n(\omega)\omega/c$ is the dispersion relation of the medium, $k_0=k(\omega_0)$ and $v_g^{-1}=k'(\omega_0)$.
The second term on the right side denotes the Fourier transform of time-dependent nonlinear response.
The nonlinear polarization includes an instantaneous Kerr response and a delayed Raman contribution with equal proportion \cite{kumar2009}:
\begin{equation}
P_{NL}=\epsilon_0n_0n_2[I+\int_{-\infty}^{t}\mathcal{R}(t-t')I(t')dt']E,
\end{equation}
where the  nonlinear refractive index coefficient $n_2=0.96\times 10^{-19}\ \mathrm{cm^2/W}$, yielding a critical power for self-focusing $P_{cr}=10$ GW \cite{liu2005a}. 
The specific form of $\mathcal{R}(t-t')$ that denotes the Raman contribution can be found in Ref. \cite{berge2007}.
The current term is composed of plasma current $J_{p}=ie^2\rho E/m_e\omega$ and ionization loss $J_{loss}=2W(|E|)U_i\rho_n/E$.
The electron density $\rho$ is calculated using the corrected Perelomov-Popov-Terent’ev formula \cite{berge2007}:
\begin{equation}
\partial_t \rho=W(|E|)\left(\rho_{\mathrm{n}}-\rho_0-\rho\right),
\end{equation}
where $W(|E|)$ is the PPT ionization rate, $U_i$ is the ionization potential of oxygen and $\rho_n$ is the neutral species density.
We assume the preformed plasma channel is uniformly distributed with transverse scale much larger than the spot size, then the effect of initial plasma density $\rho_0$ can be coupled into the dispersion relation:
\begin{equation}
 n(\omega)=n_a(\omega)-\frac{\rho_0}{2\rho_c}.
 \end{equation}
$n_a(\omega)$ is the refractive index of air \cite{zhang2008}, and $\rho_{c}=\epsilon_0m_e\omega^2/e^2$ is the frequency dependent critical plasma density that accounts for plasma dispersion.

\noindent\textbf{Spectrogram representation}
The spectrogram of the electric field is calculated as follows:
\begin{equation}
P(\tau, \omega)=\left|\int_{-\infty}^{\infty} e^{-i \omega t} E(t) h(t-\tau) d t\right|^2,
\end{equation}
where the spectrogram $P(\tau,\omega)$ is the function of time delay $\tau$ and angular frequency $\omega$.
We choose a Gaussian function with 5 fs FWHM as the gate function.
The spectrogram is calculated respectively for each $(x,y)$ data point, and then summed up.

\section{Acknowledgements}
The authors acknowledge the supports of the National Natural Science Foundation of China (NSFC) (11874056, 12074228) and Fundamental Research Funds for the Central Universities.

\section{Data availability}
The data that support the findings of this study are available from the corresponding author upon reasonable request.

\section{Author contributions}
L.X. and T.X. discussed and conceived the idea. L.X. performed the theoretical analysis and simulations. L.X. and T.X. analyzed the data and prepared the manuscript.

\section{Competing interests}
The authors declare no competing interests.

\end{document}